\documentclass[aps,pre,epsf,twocolumn,floatfix]{revtex4-2}
\usepackage{amsmath}
\usepackage{amssymb}
\usepackage{epsfig}
\usepackage{graphicx}
\usepackage[T1]{fontenc}
\usepackage[utf8]{inputenc}
\usepackage{color}
\usepackage[normalem]{ulem}
\usepackage{tikz}
\usetikzlibrary{positioning}

\begin{document}

\title{Monte Carlo study of the two-dimensional kinetic Ising model under a nonantisymmetric magnetic field}

\author{Zeynep Demir Vatansever}
\email{zeynep.demir@deu.edu.tr}

\author{Erol Vatansever}
\email{erol.vatansever@deu.edu.tr}

\affiliation{Department of Physics, Dokuz Eyl\"{u}l
University, TR-35160, Izmir-Turkey}

\author{Andreas Berger}
\email{a.berger@nanogune.eu}

\affiliation{CIC nanoGUNE BRTA, E-20018 Donostia-San Sebastian, Spain}

\author{Alexandros Vasilopoulos}
\email{alex.vasilopoulos@essex.ac.uk}

\author{Nikolaos G. Fytas}
\email{nikolaos.fytas@essex.ac.uk}

\affiliation{School of Mathematics, Statistics and Actuarial Science, University of Essex, Colchester CO4 3SQ, United Kingdom}

\date{\today}

\begin{abstract}
We present a comprehensive numerical study of dynamic phase transitions in the two-dimensional kinetic Ising model under a nonantisymmetric time-dependent magnetic field including a sinusoidal term and a second harmonic component. We demonstrate that the expected antisymmetric property and the scaling behavior of the order parameter are maintained using the recently proposed generalized conjugate field approach. Via a detailed finite-size scaling analysis we compute, for zero-bias field, the set of critical exponents suggesting that the Ising universality class is conserved, even in the absence of half-wave antisymmetry in the time-dependent magnetic field. Our results verify up-to-date experimental observations and provide a deeper understanding of non-equilibrium phase transitions, establishing a broader framework for exploring symmetry-breaking phenomena in driven magnetic systems.
\end{abstract}

\maketitle

\section{Introduction}
\label{sec:intro}

Hysteresis and dynamically ordered states, which occur in response to an oscillating control parameter, are common manifestations of non-equilibrium behavior in interacting many-particle systems~\cite{Henkel,Rao}. In systems far from equilibrium dynamically ordered states and their derivatives may undergo qualitative changes at a critical control parameter value, known as dynamic phase transition. A significant example of non-equilibrium phenomena is the dynamic phase transition observed in ferromagnetic materials exposed to a time-dependent oscillating magnetic field and it has emerged as a fundamental research field for revealing the collective dynamics behind non-equilibrium phase transitions~\cite{Chakrabarti,Yuksel}. 

Dynamic phase transitions in ferromagnets were originally observed in the Ising model subjected to a sinusoidal oscillating magnetic field, known as the kinetic Ising model~\cite{Tome}. In the kinetic Ising model, the dynamic magnetization trajectory, $M(t)$, displays significant changes at the dynamic phase transition as one modifies the external control parameter of the time-dependent field with amplitude $h_0$ and period $P$ at temperatures below the Curie temperature~\cite{Riego0}. The kinetic Ising model is characterized by the exchange interaction parameter $J$ and the intrinsic relaxation time constant, known as the metastable lifetime $\tau$. This is the time required for spins to relax to a stable state when a constant magnetic field is applied in the opposite direction to their alignment~\cite{Rikvold}. The dynamic response of the kinetic Ising model to the driving field $h(t)$ is governed by the interplay between the metastable lifetime and the field's period $P$~\cite{Sides1,Korniss}. In the slow dynamic regime, for which the period of the magnetic field is larger than $\tau$, the magnetization can reverse during the one cycle of the field, but with a phase lag, resulting in the typical hysteresis loops. This regime is referred to as the dynamic paramagnetic (disordered) phase. On the other hand, in the fast dynamic regime where the period of the magnetic field is smaller that $\tau$, the magnetization cannot follow the field. Instead, it displays small oscillations around one of the non-vanishing values of the magnetization, and this regime corresponds to the dynamic ferromagnetic (ordered) phase.

Within this framework, the average of the time-dependent magnetization over a full cycle of the periodic magnetic field
\begin{equation}\label{order_parameter}
    Q=\displaystyle \frac{1}{P} \int_t^{t+P}M(t^{\prime})dt^{\prime}
\end{equation}
is defined as the order parameter of the related dynamic phase transition~\cite{Tome}. In the case of dynamic phase transitions in ferromagnets, the system undergoes a second-order phase transition from the paramagnetic to the ferromagnetic phase at the critical period $P = P_{\rm c}$~\cite{Tome,Chakrabarti}. For $P < P_{\rm c}$, the system displays a dynamically ferromagnetic phase with $Q$ taking a non-zero value. In the dynamically paramagnetic phase ($P > P_{\rm c}$) the oscillations in the magnetization generate a vanishing value of the order parameter with $Q = 0$. 

Over the years, many concepts and tools from the pool of thermodynamic phase transitions have been applied for the study of dynamic phase transitions in ferromagnets. These include experimental work~\cite{Le, Jiang, Suen, Kleemann, Robb0, Berger, Quintana0}, theoretical approximations (mainly of mean-field type)~\cite{Zimmer,Buendia1,Jang1,Shi,Punya,Gallardo,Riego}, but also Monte Carlo simulations~\cite{Korniss, Lo, Acharyya1, Acharyya2, Buendia2, Acharyya3, Buendia4}. The above vast literature has revealed that conventional thermodynamic and dynamic phase transitions share many analogies~\cite{Riego0}. In particular, both are described by relevant order parameters ($M$ and $Q$) that manifest a continuous phase transition from ferromagnetic to paramagnetic phases at critical points, $T_{\rm c}$ and $P_{\rm c}$, respectively, and have similar phase diagrams. In the last two decades, significant efforts have been made to estimate critical exponents and the underlying universality classes of the kinetic Ising model and its variants. Applications of finite-size scaling techniques to the kinetic Ising model have demonstrated that thermodynamic and dynamic phase transitions belong to the same universality class~\cite{Sides1,Sides2,Korniss,Buendia3,Vatansever2, Park2,Vatansever_Fytas,Vatansever3,Vasilopoulos,Vatansever4}. It is worth noting that these findings also support the symmetry arguments put forward by Grinstein \emph{et al.}~\cite{Grinstein} and the study of a Ginzburg-Landau model in an external oscillating field~\cite{Fujisaka}.  Furthermore, theoretical results have been validated through a recent experimental investigation on dynamic critical properties of ultrathin uniaxial Co films, showing that the dynamic critical exponents are consistent with those predicted for the two-dimensional Ising ferromagnet~\cite{Quintana0}.

Additional analogies between thermodynamic and dynamic phase transitions emerge when a ferromagnet is exposed to a time-independent constant-bias field, $h_{\rm b}$, applied in addition to the periodic field. It has been shown that the bias field serves as the conjugate field of the order parameter, similar to the external magnetic field $h$ being the conjugate field of $M$ in the equilibrium set up~\cite{Gallardo,Robb1}. This latter observation has stimulated the investigation of the order parameter in two-dimensional  $P-h_{\rm b}$ and $h_0-h_{\rm b}$ dynamic phase spaces in several studies~\cite{Berger, Quintana0, Ramirez, Quintana} yielding new insights into various aspects of the relevant phase diagrams. One of these is the time-reversal symmetry in the dynamic phase diagram, \emph{i.e.}, 
\begin{equation}\label{antiQ}
    Q(h_{\rm b}) = -Q(-h_{\rm b})
\end{equation}
which is valid for all $P$ and $h_0$ values in the dynamic phase space~\cite{Robb,Quintana1,Quintana2}. This antisymmetric property of the order parameter applies to the time-dependent part of the magnetic field sequences that display half-wave antisymmetry, namely $h(t)=-h(t+t_{1/2})$, where $t_{1/2}$ is the half-period value of the external field. In the dynamically paramagnetic phase, the half-wave antisymmetry of $h(t)$ causes the magnetization curve to exhibit antisymmetric behavior, with $M(t) = -M(t + t_{1/2})$, leading to a vanishing order parameter ($Q = 0$). To date, the time-dependent field applied to magnetic systems is typically chosen as sinusoidal in experiments and mean-field studies, while square-like fields are used in Monte Carlo simulations, both exhibiting the half-wave antisymmetric property. A nonantisymmetric magnetic field with the lack of half-wave antisymmetry can be expressed as~\cite{Quintana1, Quintana2} follows
\begin{equation}\label{field}
    h(t)=h_{\rm b}+h_0 \sin \left( \frac{2 \pi t}{P} \right) +h_2\sin  \left( \frac{4 \pi t}{P} \right).
\end{equation}
Here, the first and second terms denote the bias field and the sinusoidal magnetic field component with amplitude $h_0$ and period $P$, respectively. The last term is the second harmonic contribution of period $P/2$ and amplitude $h_2$. This component is the lowest-order even Fourier component and is responsible for the broken-time antisymmetry of the field. 

Recent experimental and theoretical observations within the mean-field approximation indicated that in the presence of fields with a lack of half-wave antisymmetry, $M(t)$ is no longer antisymmetric when $h_{\rm b} = 0$, resulting in $Q \ne 0$ in the dynamically paramagnetic phase~\cite{Quintana1,Quintana2}. Moreover, deviations from the antisymmetric behavior of $Q$ as a function of $h_{\rm b}$ have been observed to become more pronounced with increasing $h_2$~\cite{Quintana2}. According to these findings, when the half-wave antisymmetry of the external field is lost, the constant-bias field $h_{\rm b}$ is no longer a conjugate field, and a general definition of the conjugate field $h^{\ast}$, that preserves the time-reversal symmetry of the dynamic phase diagram, is needed. In recent studies, a general definition of the conjugate field is proposed as~\cite{Quintana1, Quintana2}
\begin{equation}\label{conjugate_field}
h^{\ast}=h_{\rm b}+\Delta h
\end{equation}
where $\Delta h$ is the nonlinear effective bias correction
\begin{equation}\label{conjugate_field2}
\Delta h = \displaystyle -\frac{1}{2} \left[ h_{\rm b}(Q) +h_{\rm b}(-Q) \right].
\end{equation}
It has been shown both experimentally and within the mean-field theory that the definition of $h^{\ast}$ in Eq.~\eqref{conjugate_field} recovers the expected antisymmetry of the order parameter as a function $h_{\rm b}$~\cite{Quintana1,Quintana2} as
\begin{equation}\label{antiQstar}
    Q(h^{\ast}) = -Q(-h^{\ast}).
\end{equation}
According to mean-field calculations conducted near the dynamic phase transition in the presence of periodic fields that lack half-wave antisymmetry, the scaling behavior of the dynamic order parameter is conserved if one uses the general definition of the conjugate field defined in Eq.~\eqref{conjugate_field} above. Furthermore, the extracted critical exponents of the order parameter have been found to agree with those predicted by the mean-field model even for large amplitudes of $h_2$, thereby confirming the concept of universality.

At this stage, further investigation is required to validate the general definition of $h^{\ast}$ as the true conjugate field in dynamic phase transitions using other methods where several crucial factors, such as the spin-spin correlations, and the dimensionality/topology of the lattice can be taken into account. In this respect, we present in the current work an extensive Monte Carlo investigation of the dynamic magnetic phase diagram of the square-lattice kinetic Ising model in the presence of a nonantisymmetric field sequence $h(t)$ that includes a fundamental harmonic component $h_0$ and a second-harmonic component $h_2$. A detailed finite-size scaling analysis based on several thermodynamic quantities allows us to explore the dynamic phase diagram and to scrutinize the universality aspects. Alongside our main analysis, we also extract the critical exponent $\delta$ of $Q(h^{\ast})$ at the dynamic phase transition for various values of $h_2$ and show that the generalized conjugate field $h^{\ast}$ is indeed the suitable conjugate field of the dynamic order parameter.

The outline of the remaining parts of the paper is as
follows: In Sec.~\ref{sec:framework} we introduce the model and the details of our simulation protocol. We also define the relevant observables that will facilitate our finite-size scaling analysis for the characterization of the universality principles of this dynamic phase transition. The numerical results and discussion are presented in Sec.~\ref{sec:results}. Finally, Sec.~\ref{sec:conclusions} presents a summary of our conclusions.

\section{Simulation framework}
\label{sec:framework}

\subsection{Model and numerical details}
\label{sec:model}

In this work we considered the fruit-fly model of Statistical Physics, \emph{i.e.}, the two-dimensional square-lattice kinetic Ising model under a nonantisymmetric magnetic field sequence. The Hamiltonian of the system reads as
\begin{equation}\label{Hamiltonian}
\mathcal{H} = -J\sum_{\langle x y \rangle} \sigma_x \sigma_y - h(t) \sum_{x}{\sigma_x},
\end{equation}
where $\sigma_{x} = \{ \pm 1 \}$ is the spin variable , $\langle xy \rangle$ indicates summation over nearest neighbors, and $J>0$ denotes the ferromagnetic exchange interaction. The final term is the Zeeman energy, with $h(t)$ representing a spatially uniform, periodically oscillating magnetic field; see Eq.~\eqref{field}. 

In numerical grounds, we carried out Monte Carlo simulations on square lattices with periodic boundary conditions
using the single-site update Metropolis algorithm~\cite{Metropolis,Binder,Newman,Binder2}. In our simulations, $N = L \times L$ defines the total number of
spins and $L$ the linear dimension of the lattice, taking values within the range $L = \{24 - 1024\}$. To facilitate the numerical process, we implemented a geometric parallelization procedure where the lattice is divided into strips of $L \times L/N_{\rm p}$, with $N_{\rm p}$ the number of available processors. For each set of simulation parameters we performed $100$ independent computer experiments (allowing thus the computation of errors via the jackknife method~\cite{Newman}), using the following protocol: the first $10^3$ periods of the external field have been discarded during the thermalization process and numerical data were collected and analyzed during the following $10^4$ periods of the field. In our simulations the unit of time is defined as one Monte Carlo step per site (MCSS). Note also that our simulations were performed in the multi-droplet regime, in which the decomposition of the metastable phase arises via the nucleation and growth of numerous droplets of the stable phase~\cite{Rikvold,Sides1}. It is well established that the metastable decay of the system during field reversals is influenced by the temperature, field strength, and system size~\cite{Korniss}. Thus, to examine the system in this multi-droplet regime, we have set the amplitude of the external magnetic field to $h_{0} = 0.3$ and fixed the temperature at $T = 0.8T_{\rm c}$~\cite{Park2}, where $T_{\rm c}$ is the Curie temperature of the equilibrium square-lattice Ising model~\cite{Onsager}. 

A comment for the application of finite-size scaling to the numerical data: We have restricted ourselves to data with $L\geq L_{\rm min}$, adopting the standard $\chi^{2}$ test for goodness of the fit. Specifically, we considered a fit as being acceptable only if $10\% < p < 90\%$, where $p$ is the quality-of-fit parameter~\cite{Press92}.

\subsection{Observables}
\label{sec:obs}

The main quantity of interest is the time-dependent magnetization per site
\begin{equation}\label{eq:2}
 M(t)=\frac{1}{N}\sum_{x = 1}^{N}\sigma_{x}(t),
\end{equation} 
which, when integrated over one cycle of the field, see Eq.~\eqref{order_parameter},
provides access to the dynamic order parameter $Q$. Given that the probability density of the order parameter exhibits two opposite peaks in such finite systems, we measure the average norm of the order parameter, $\langle |Q| \rangle$, in the calculations. Furthermore, to identify and characterize dynamic phase transitions and to extract critical exponents using finite-size scaling methods, it is essential to compute the scaled variance of the dynamic order parameter
\begin{equation}\label{eq:4}
\chi = N\left[\langle Q^2\rangle -\langle |Q|
\rangle^2 \right],
\end{equation}
which is analogous to the static susceptibility~\cite{Sides1, Sides2, Korniss}. The use of $\chi$ has been validated as a proxy for nonequilibrium susceptibility through fluctuation-dissipation relations~\cite{Robb1}. Finally, with the help of the dynamic order parameter $Q$ we
may define the corresponding fourth-order Binder cumulant
\begin{equation}\label{eq:6}
 U_4=1-\frac{\langle Q^4\rangle}{3\langle Q^2
\rangle^2},
\end{equation}
which provides us with an alternative estimation of the critical point, at the same giving time a flavor of universality at its intersection point~\cite{Binder81}.
\begin{figure}[h!]
\begin{center}
\includegraphics[scale=1.0]{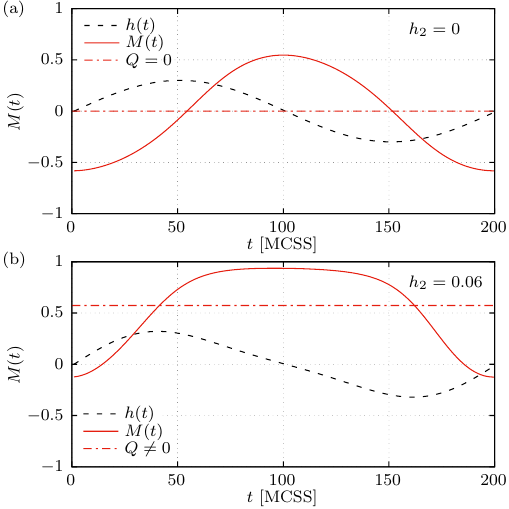}
\caption{Time evolution of the magnetization $M(t)$ (red solid curves) of the square-lattice kinetic Ising model under a time-dependent magnetic field defined in Eq.~\eqref{field} (black dashed lines). Panel (a) shows the case with zero second harmonic component ($h_2 = 0$) where the magnetization exhibits antisymmetric behavior, resulting in a dynamic order parameter of $Q = 0$. Panel (b) showcases the scenario with $h_2 = 0.06$, breaking the half-wave antisymmetry and leading to a non-zero dynamic order parameter ($Q \neq 0$). Both plots correspond to a lattice size of $L = 1024$ and a half-period of $t_{1/2} = 100$ MCSS. The numerical data shown here are obtained by averaging over $10^{4}$ periods of the external field.} \label{Fig1}
\end{center}
\end{figure}

\section{Results and discussion}
\label{sec:results}

We start the presentation of our results with Fig.~\ref{Fig1}, displaying the time evolution of the magnetization for two selected values of $h_2$ in the dynamic paramagnetic phase. To facilitate the discussion we note here that the critical half-period of the system is around $93$ MCSS (see also Fig.~\ref{Fig4}(a) below) . When $h_2=0$ (panel (a)), one obtains a typical behavior observed previously in the kinetic Ising model, where the magnetization can follow the field with a phase lag~\cite{Korniss,Vatansever3}. Since the time-dependent magnetic field has the property of half-wave antisymmetry, the magnetization has an antisymmetric behavior with time leading to a period-averaged value of $Q = 0$. However, when the half-wave antisymmetry of $h(t)$ is broken due to the existence of the second-order Fourier component of the field with an amplitude of $h_2=0.06$ (panel (b)), the antisymmetric behavior of the magnetization also disappears. This leads to a finite dynamic order parameter $Q \ne 0$, despite the magnetization undergoing a full reversal cycle as observed in recent experimental and theoretical studies~\cite{Quintana1,Quintana2}. 

\begin{figure*}[h!]
\begin{center}
\includegraphics[scale=1.0]{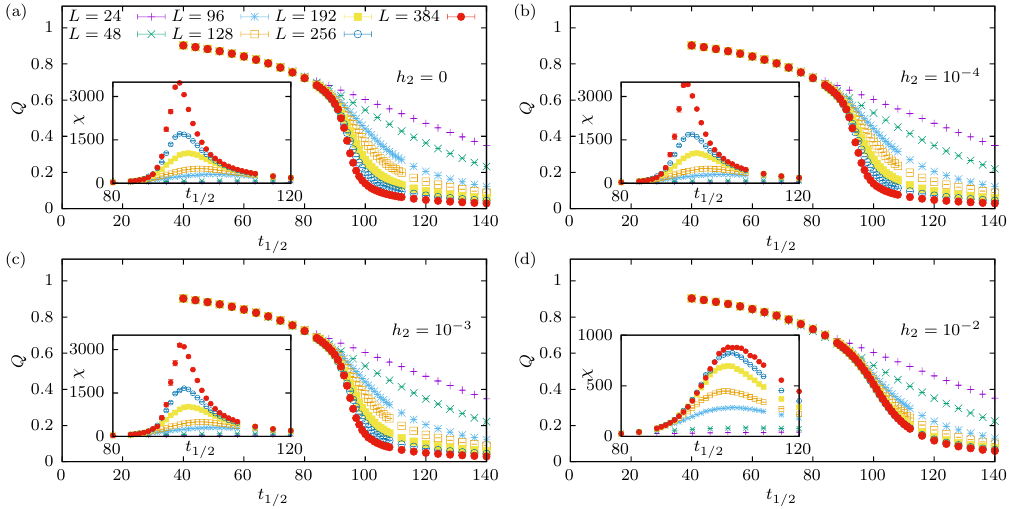}
\caption{Half-period dependence of the dynamic order parameter $Q$ (main panels) and magnetic susceptibility $\chi$ (insets) for various lattice sizes and different values of the second harmonic component $h_2$. Panels (a) to (d) correspond to $h_2 = 0$, $10^{-4} $, $10^{-3}$, and $10^{-2} $, respectively. For small values of $h_2$, panels (a)–(c), the order parameter $Q$ exhibits a clear second-order phase transition with strong finite-size effects near the dynamic phase transition point. As $h_2$ increases the height of the magnetic susceptibility peak diminishes, and the sharp phase transition disappears at $h_2 = 10^{-2}$, as shown in panel (d). The insets illustrate the divergence of $\chi$ with increasing system size $L$, characteristic of second-order phase transitions, at least in panels (a)-(c). Clearly, no such divergence is observed in panel (d).}\label{Fig2}
\end{center}
\end{figure*}

\begin{figure*}[h!]
\begin{center}
\includegraphics[scale=1.0]{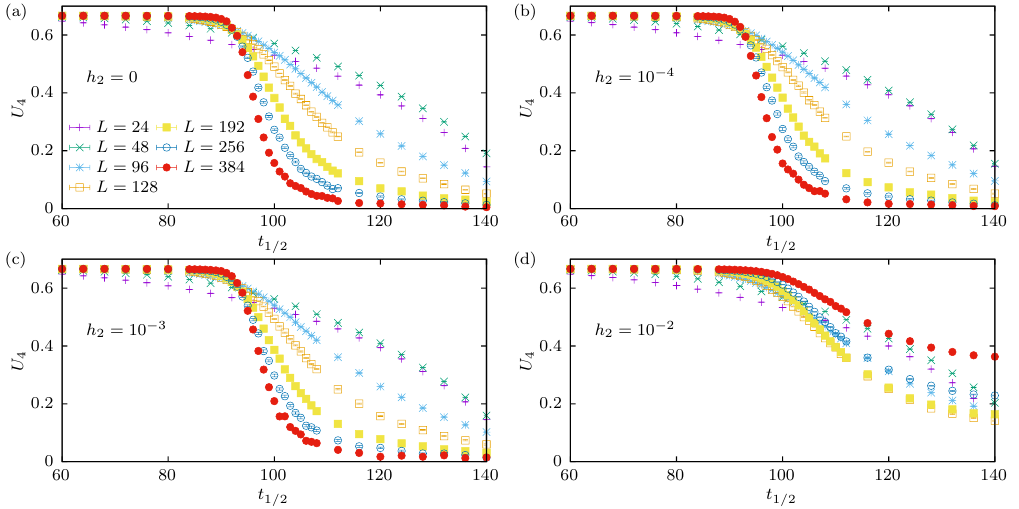}
\caption{Half-period dependence of the fourth-order Binder cumulant $U_4$. The simulation parameters (\emph{i.e.}, $L$ and $h_2$ values) are identical to those of Fig.~\ref{Fig2}. A clear crossing behavior can be observed for all values of $h_2 < 10^{-2}$ (panels (a)-(c)), except for the data corresponding to $h_2 = 10^{-2}$ (panel (d)), where no crossing is observed, indicating the absence of a dynamic phase transition.}\label{Fig3}
\end{center}
\end{figure*}

To understand the overall behavior of the order parameter across a broad dynamic phase space, we now examine the variation of $Q$ with the half-period of the external field for several values of $h_2$ and a wide range of system sizes, as illustrated in the main panel of Fig.~\ref{Fig2}. In the dynamic ferromagnetic phase, the half-period dependence of $Q$ exhibits a typical behavior, showing a finite value and no observable finite-size effects
for all values of $h_2$. Also, the presence of the $h_2$  term in the $h(t)$ formula does not cause any obvious modification in the value of $Q$ and the dynamic phase transition point which will be discussed later in our finite-size scaling analysis. In the dynamic paramagnetic phase, and the absence of $h_2$, $Q = 0$ is expected, as mentioned also while inspecting Fig.~\ref{Fig1}. It is evident from Fig.~\ref{Fig2} that there are strong finite-size effects for small $L$-values, which weaken upon increasing the system size. If one concentrates on the largest lattice size of $L = 384$ shown, there is a slight increment in the value of $Q$ for larger values of the half period which is particularly noticeable for $h_2 = 10^{-2}$. For small values of $h_2$ (namely $h_2=0$, $10^{-4}$, and $10^{-3}$ considered here), the order parameter exhibits a pronounced change near the dynamic phase transition corresponding to the peak observed in the dynamic susceptibility, $\chi$, given in the inset of Fig.~\ref{Fig2}. Additionally, the dynamic susceptibility $\chi$ exhibits a clear divergence, with its height increasing as $L$ takes larger values, indicating a second-order phase transition.
On the other hand, this divergent behavior of $\chi$ is not present at the relatively larger value of $h_2=10^{-2}$, as indicated by the smooth variation of $Q$ with $t_{1/2}$. This behavior is also a signature of the absence of a dynamic phase transition for the larger value of $h_2$ considered, \emph{i.e.}, $h_2=10^{-2}$. 

As already observed in experiments and theoretical studies~\cite{Quintana1,Quintana2}, introducing the second harmonic field component breaks the half-wave antisymmetric property of $h(t)$, resulting in an effective bias effect. Since this bias effect increases with $h_2$, one expects that there should be no dynamic phase transition for relatively large values of $h_2$. To demonstrate this, we use the fourth-order Binder cumulant of the order parameter, $U_4$, which provides critical information about the existence of a phase transition. The half-period dependency of the Binder cumulant is shown in Fig.~\ref{Fig3} for the same set of parameters as those used in Fig.~\ref{Fig2}. For all values of $h_2 < 10^{-2}$, the $L$-dependent cumulant curves intersect, marking on the horizontal axis the critical half-period period of the system and at the vertical axis verifying (within error bars) the universal value $U^{\ast} = 0.610\:692\:4(16)$ of the two-dimensional equilibrium Ising model~\cite{Salas}. Certainly, the crossing point is expected to depend on the lattice size $L$ (as also shown in the figure) and the term universal is valid for given lattice shapes, boundary conditions, and isotropic interactions. For a detailed discussion of this topic we refer the reader to Refs.~\cite{Selke1} and \cite{Selke2}. Still, the scope of the current Fig.~\ref{Fig3} is to show qualitatively an instance of the Ising universality. Note however that there is no crossing for the case with $h_2=10^{-2}$ supporting the above discussion regarding the absence of a dynamic phase transition for this particular case.   
\begin{figure}[h!]
\begin{center}
\includegraphics[scale=1.0]{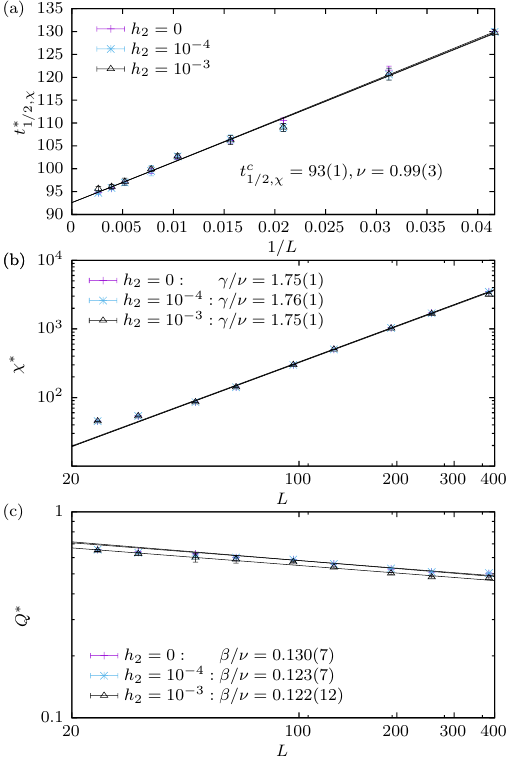}
\caption{(a) Determination of the critical half-period $t_{1/2,\chi}^{\rm c}$ and the correlation-length's critical exponent $ \nu $ using finite-size scaling analysis. The shift behavior of the pseudocritical half-period $t^{\ast}_{1/2,\chi} $ is plotted as a function of $1/L$, with the solid lines representing a simultaneous fit to the power-law ansatz, Eq.~\eqref{tc}. (b) Finite-size scaling of the peak of the dynamic susceptibility $\chi^{\ast}$ for several values of $h_2$. The solid lines show fits according to Eq.~\eqref{chi}. (c) Finite-size scaling of the dynamic order parameter $Q^{\ast}$ at the critical half-period, following Eq.~\eqref{beta}. Note the double-logarithmic scale in panels (b) and (c).}\label{Fig4}
\end{center}
\end{figure}

Having determined the values of $h_2$ at which a dynamic phase transition occurs, we employ now the standard finite-size scaling analysis to explore the universality principle of the model and corroborate the quantitative results of Fig.~\ref{Fig3} based on the crossings of the Binder cumulant. It should be noted that earlier works on the kinetic Ising and Blume-Capel models and their variants have shown that the finite-size scaling laws can also be effectively applied to systems far from equilibrium~\cite{Korniss, Vatansever2, Vatansever_Fytas, Vatansever3, Vasilopoulos, Vatansever4}. 
\begin{figure}[h!]
\begin{center}
\includegraphics[scale=1.0]{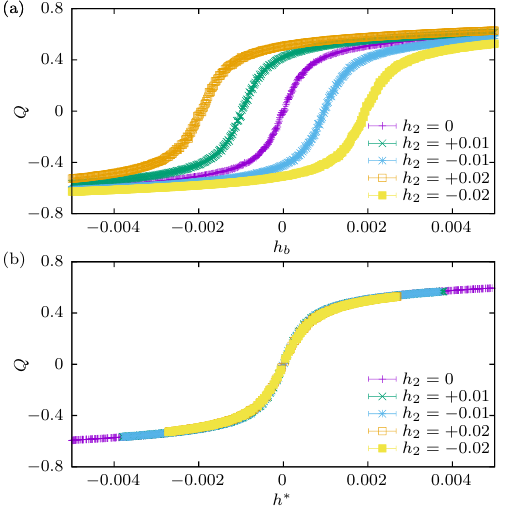}
\caption{Dynamic order parameter $Q$ as a function of the bias field $h_{\rm b}$ (a) and the generalized conjugate field $h^{\ast}$ (b), for several values of $h_2$ in the dynamic paramagnetic phase ($t_{1/2} = 100$ MCSS). The numerical data shown correspond to a system with linear size of $L = 256$.}\label{Fig5}
\end{center}
\end{figure}
The location of the maxima of the dynamic susceptibility in Fig.~\ref{Fig2} can be referred to as suitable pseudocritical half-periods, denoted hereafter as $t^{\ast}_{1/2,\chi}$. The shift behavior of these pseudocritical half-periods is plotted in Fig.~\ref{Fig4}(a). The solid lines highlight a simultaneous fit of 
the form~\cite{Fisher,Privman,Binder92}
\begin{equation}\label{tc}
 t^{\ast}_{1/2,\chi}= t^c_{1/2,\chi} +b L^{-1/\nu},
\end{equation}
where the critical half-period $t^{\rm c}_{1/2,\chi}$ and the critical exponent $\nu$ are shared among the three data sets, and $b$ is a non-universal fitting constant. 
The obtained value $t^c_{1/2,\chi} = 93(1)$ MCSS for the critical half-period is compatible with the location of the crossing points of the Binder cumulants identified in Fig.~\ref{Fig3} and the estimate $\nu = 0.99(3)$ for the correlation-length exponent points again to the universality class of the equilibrium two-dimensional Ising model. Subsequently, in Fig.~\ref{Fig4}(b) we present the finite-size scaling behavior of the maxima of the dynamic susceptibility, $\chi^{\ast}$, again for the lower $h_2$ values studied. Here, the solid lines represent fits of the form
\begin{equation}\label{chi}
    \chi^{\ast} \sim L^{\gamma/\nu},
\end{equation}
giving exponent ratios in excellent agreement with the exact result $\gamma/\nu = 7/4$ of the Ising ferromagnet, as outlined also in the panel.
In addition to $\gamma/\nu $, further evidence may be provided via the alternative magnetic exponent ratio, namely, $\beta/\nu$, obtained from the scaling behavior of the dynamic order parameter at
the critical point via
\begin{equation}\label{beta}
    Q^{\ast} \sim L^{-\beta/\nu}.
\end{equation}
\begin{figure}[h!]
\begin{center}
\includegraphics[scale=1.0]{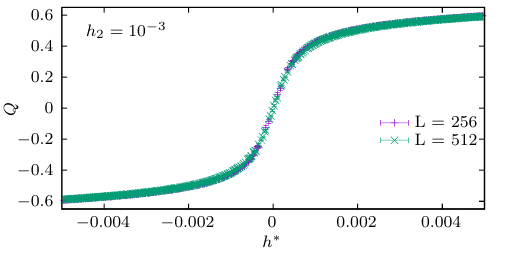}
\caption{Dynamic order parameter $Q$ as a function of the generalized conjugate field $h^{\ast}$ for two different lattice sizes, $L = 256$ and $L = 512$, with a second harmonic field component $ h_2 = 10^{-3} $ in the dynamic paramagnetic phase.}\label{Fig6}
\end{center}
\end{figure}
In Fig.~\ref{Fig4}(b) we depict this scaling behavior, where power-law fits of the form~\eqref{beta} give estimates for $\beta/\nu$ compatible with the Ising value $1/8 = 0.125$. A summary of estimates for the critical-exponent ratios $\gamma/\nu$ and $\beta/\nu$ for non-vanishing, but sufficiently low values of the second harmonic field component $h_2$ are given in Table~\ref{table}. At this point we would like to note that we also included in our fits scaling corrections (in the form of polynomials in $\sim L^{-\omega}$) where $\omega = 1.75$, see, for example, the discussion in the supplementary material of Ref.~\cite{Shao}. We note, however, that other values of $\omega$ have also been reported for certain quantities in the two-dimensional Ising model, most notably $\omega = 4/3$ and $\omega = 2$, and in some cases also the analytic corrections might be dominant. In our case, the observed corrections were so weak (compared to the statistical accuracy of our data) that numerically we could not reliably identify any difference between these choices and even by excluding the correction terms.

So far our discussion was restricted to the case with zero-bias field, indicating that the (dynamic) ferromagnetic phase is largely unaffected by the presence of $h_2$, as long as its value is sufficiently low. Clearly, the next indispensable step forward refers to the application of a time-independent bias field at, and above, the critical half-period, targeting at a deeper understanding of the effect of $h_2$ on the order-parameter's critical behavior. In Fig.~\ref{Fig5}(a), we show $Q$ as a function of $h_{\rm b}$ for various values of $h_2$ in the dynamic paramagnetic phase phase ($t_{1/2}=100$ MCSS). As one can observe, in the absence of $h_2$ the inversion of $h_b \rightarrow - h_b$ in the magnetic field leads to exactly $Q \rightarrow - Q$, given in Eq.~\eqref{antiQ}. When $h_2$ is non-zero, deviations from the antisymmetric property become apparent and more pronounced with increasing $h_2$. If both $h_2$ and $h_b$ are nonzero, the $Q \rightarrow - Q$ state is accessed with $h_2 \rightarrow - h_2$ and $h_b \rightarrow - h_b$, \emph{i.e.}, $Q(h_2,h_{\rm b})=-Q(-h_2,-h_{\rm b})$.
\begin{figure}[h!]
\begin{center}
\includegraphics[scale=1.0]{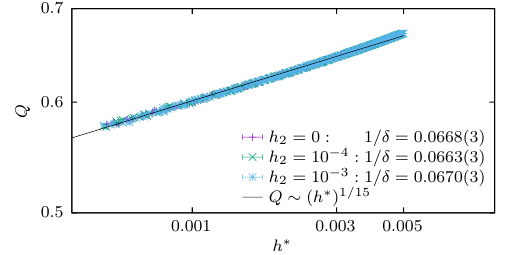}
\caption{Dynamic order parameter $Q$ as a function of the generalized conjugate field $h^{\ast}$ at the critical half-period for different values of the second harmonic field component $h_2$ for a system with linear size $L = 256$. The black solid line represents the scaling behavior $Q\sim (h^{\ast})^{1/\delta} $, with $ \delta = 15 $. Note the double-logarithmic scale of the panel.}\label{Fig7}
\end{center}
\end{figure}
It is clear that when $h_2$ is added to the standard magnetic field sequence, $h_{\rm b}$ no longer accurately represents the conjugate field. Accordingly, as the next step, we utilize the general definition of the conjugate field, $h^{\ast}$, introduced very recently by Quintana and Berger~\cite{Quintana1,Quintana2} to restore the antisymmetric behavior of the order parameter. Figure~\ref{Fig5}(b), displays $Q$ vs $h^{\ast}$, as computed from the data of Fig.~\ref{Fig5}(a) by using Eqs.~\eqref{conjugate_field} and \eqref{conjugate_field2}. Here, Monte Carlo simulation results indicate that for the considered amplitudes of $h_2$, the antisymmetry of the $Q-h^{\ast}$ curve is recovered with all the curves nearly overlapping. This result shows that the proposed $h^{\ast}$ is the true conjugate field and confirms the earlier experimental and theoretical investigations~\cite{Quintana1, Quintana2}. 

To explore the existence of possible finite-size effects in the above discussion, we present in Fig.~\ref{Fig6} two characteristic $Q-h^{\ast}$ curves for a selected value of $h_2=10^{-3}$ in the dynamic paramagnetic phase and two system sizes, namely $L=256$ and $L=512$. It is evident from this figure that there are no signs for significant finite-size effects (in fact the curves for the two different $L$'s practically overlap) for the considered value of $h_2$. Hence, in the following analysis focusing on the numerical determination of the critical exponent $\delta$ we shall restrict our analysis on the $L = 256$ system size.

At the critical half-period value of the external field the order parameter is known to display scaling behavior of the form $Q \sim (h_{\rm b})^{1/\delta}$, where $\delta=1+\gamma/\beta$~\cite{Robb1,Quintana2}.
In full analogy with the equilibrium case, the relevant exponent in the kinetic Ising model was determined in previous studies (where half-wave antisymmetry of $h(t)$ was present) to take the value $\delta = 15$. Since it has been demonstrated previously within mean-field approximation and in our above analysis that $h^{\ast}$ is the true conjugate field that establishes the antisymmetric property of $Q$, it is possible to write at the critical period an analogous ansatz~\cite{Quintana1,Quintana2}
\begin{equation}\label{delta}
    Q \sim (h^{\ast})^{1/\delta}.
\end{equation}
In Fig.~\ref{Fig7}, we sketch the dynamic order parameter $Q$ vs. $h^{\ast}$ for different $h_2$ values for a system with linear size $L=256$. The extracted exponents obtained from the slope of the $Q-h^{\ast}$ curves are also listed in Table~\ref{table} and they are very close to the exact value $1/\delta = 1/15 \approx 0.067$. Consequently, our numerical data and analysis confirm that the scaling property of Eq.~\eqref{delta} is valid for the considered values of $h_2$ and the generalized field defined in Eq.~\eqref{conjugate_field} and also that Eq.~\eqref{conjugate_field2} can be used as a conjugate field when half-wave antisymmetry is lost in the time-dependent magnetic-field term. 

\begin{table}
	\caption{A summary of critical-exponent ratios $\gamma/\nu$, $\beta/\nu$, and $1/\delta$ of the dynamic susceptibility $\chi$, dynamic order parameter $Q$, and generalized conjugate field $h^{\ast}$ for various values of the second harmonic field component $h_2$, as computed in the present work. The last row of the table provides for reference the exact values marking the universality class of the equilibrium two-dimensional Ising ferromagnet.}\label{table}
\begin{ruledtabular}
\begin{tabular}{lccc}     
			$h_2$ &  $\gamma/\nu$ & $\beta/\nu$  & $1/\delta$ \\
				\hline
			$0$  & $1.75(1)$ &$ 0.130(7)$ & $0.0668(3)$\\
			$10^{-4}$  & $1.76(1)$ & $0.123(7)$  & $0.0663(3)$\\
			$10^{-3}$ &  $1.75(1)$ & $0.122(4)$  & $0.0670(3)$  \\ \hline \hline 
 Exact values &  $7/4$ & $1/8$  & $1/15$ \\
\end{tabular}
\end{ruledtabular}
\end{table}

\section{Conclusions}
\label{sec:conclusions}

In the present manuscript we have investigated the dynamic critical behavior of the two-dimensional kinetic Ising model under a nonantisymmetric magnetic field, using extensive Monte Carlo simulations and finite-size scaling techniques. By introducing a second harmonic component to the magnetic field which breaks the traditional half-wave antisymmetry, we observed significant modifications in the underlying dynamic phase transitions.
We confirmed that the generalized conjugate field restores the expected antisymmetry in the dynamic order parameter, in agreement with recent experimental findings. Our finite-size scaling analysis showed that, despite the broken half-wave antisymmetry, the universality class and critical exponents of the model remain consistent with those of the equilibrium two-dimensional Ising model. Further investigation is needed in this direction for larger values of the amplitude of the second harmonic field component in order to scrutinize the requirement of conjugate-field corrections, as well as the universality properties of spin models driven by a time-dependent magnetic field. These results validate the theoretical predictions on symmetry-breaking phenomena and extend our understanding of non-equilibrium systems subjected to nonantisymmetric fields. Moreover, our study confirms that the critical behavior of the kinetic Ising model is robust, even in the presence of symmetry-breaking perturbations, offering direct theoretical support for recent experimental observations. These findings provide a comprehensive framework for studying dynamic phase transitions in driven magnetic systems and pave the way for further experimental and theoretical work in the exploration of symmetry-breaking effects in complex systems.

\begin{acknowledgments}
The numerical calculations reported in this paper were performed at T\"{U}B\.{I}TAK ULAKBIM (Turkish agency), High Performance and Grid Computing Center (TRUBA Resources). The work of AV and NGF was supported by the  Engineering and Physical Sciences Research Council (grant EP/X026116/1 is acknowledged). Work at nanoGUNE was supported by the Spanish Ministry of Science and Innovation under the Maria de Maeztu Units of Excellence Program (Grant No. CEX2020-001038-M) and Project No. PID2021-123943NB-I00 (OPTOMETAMAG).
\end{acknowledgments}

\end{document}